\documentclass{article}

\begin{document}

\title{Unidentified $\gamma$-ray sources: new source classes
      with GLAST}


\author{Diego F. Torres \\
ICREA and Institut de Cienci\`es de l'Espai \\ IEEC-CSIC, E-08193 Barcelona, Spain. \\
dtorres@ieec.uab.es}



\maketitle

\begin{abstract}

New source classes are expected to appear in the GLAST/LAT Catalog. Here, the problems faced for their identification are summarized, and some key features of the most likely new populations of the $\gamma$-ray sky are mentioned.

\end{abstract}



The EGRET sources appear to arise from several populations, although just two classes have been identified. Firm evidence favoring the positive identification of blazars comes from the simultaneous detection of variability in multifrequency observations of key individuals, e.g., 3C 279. The second class is formed by pulsars. Here again, firm evidence is available: the detection of pulsations with the charateristic frequency in $\gamma$-rays.
It is not by chance that the two  populations detected in EGRET data present variability or periodicity, a time signature that can be tracked in other frequencies.   EGRET has had a limited power for the localization of sources (typical positional uncertainty around 1 deg), accordingly, IDs were achieved for the
brightest sources with smallest positional uncertainties, and the vast
majority outside the Galactic plane avoiding the problems induced
from dominant galactic diffuse emission. Thus, although statistical claims can be made for many other Galactic source classes (e.g., \cite{rbt}) an unambiguous detection of an individual to them pertaining has not been achieved.
EGRET has also detected the LMC, albeit not defined a population along with it (the LMC was the only galaxy other than the Milky Way detected in its diffuse emission). In this case, it was not variability, but extension, what gave confidence on the reliability of the identification. 
All in all, EGRET left us with hints of several new $\gamma$-ray populations underneath the identifications achieved. We expect new classes of sources 
mainly by two  reasons: 
{\it a) Phenomenological reasons:}
-There is a low chance probability for spatial coincidences between EGRET sources \& members of new classes, even accepting the caveats of non-uniform EGRET coverage and over subscription of possible counterparts candidates (a SNR and a stellar association being close in the sky, both superposed to the same source).
 -There are $\gamma$-ray features that are hard to encompass within detected populations, like those associated with variability behavior beyond the known source classes, and the known distributions in the sky.
 -There are a significant number of sources presenting extension \& confusion (even admitting composite sources), both beyond the appearance of known classes. 
 -There are FoM approaches pointing to $\gamma$-ray sources orphan from known counterparts, despite dedicated searches.
 -There are new populations in nearby energy bands, especially at higher energies (e.g., H.E.S.S. - MAGIC recent discoveries of new $\gamma$-ray source classes, like $\gamma$-ray binaries and pulsar wind nebulae).
{\it  b) Theoretical expectations:}
 -The expected $\gamma$-ray output for different objects have been computed in detail, and it is above the sensitivity of EGRET (for some cases) and LAT (for many more cases).

How well will GLAST be able to identify the individuals and the classes as a whole? Does one thing depend on the other? How? What kind of problems will arise?

\section{Identification of sources in perspective}

Four kinds of problems related to source identification can be distinguished:

 {\it The problem of the sheer number of sources and possible counterparts: }{\small 
An overlay of all error boxes of GRBs reported from BATSE
covers the whole sky. Consequently,
a spatial correlation analysis lacks identification capability. 
If LAT detects 1000 
sources at galactic latitudes less than 10 degrees, each with a positional uncertainty of 
12 arcmin, 20\% of the relevant sky is covered. If LAT detects 10000 sources, out of this latitude range, each with 30 arcmin uncertainty, again, 20\% of the relevant sky is covered. }

 {\it The problem of going deep in a not-sufficiently small localization error:}{\small 
 Consider the case of an average EGRET source recently studied \cite{la}: 3EG J1249-8330
(for which the error localization is 0.66 deg, and the $\gamma$-ray flux is $ 2 \times 10^{-7}$ ph cm$^{-2}$ s$^{-1}$. After 4 XMM-EPIC pointings, 148 X-ray sources were detected. Clearly, an statistical evaluation of counterparts must proceed. But does computing a redefined FoM, something like $p(i)_{c} =p_{pos} \times p(i)_{SED} \times p(i)_{var} \times p(i)_{ext} \times \ldots $, where $p(i)_{c}$ is the probability for source $i$ to be the counterpart, and each right factors represents the contribution to this  probability related to position, energy distribution, variability, extension, etc., yield a source identification here? It is not hard to convince oneself that the answer is most likely not, since for such number of  counterparts candidates, the probability $p(i)_c$ for many sources will be undistinguishable already at the systematics of its computation. }

 {\it  The problem of concurrent spatial coincidences with counterparts pertaining to different classes, even for well localized sources: }{\small 
Here take as an example the source detected at high energies by H.E.S.S., J1303-631, whose positional determination is much beyond the typical EGRET sources,  (RA=13h 03m 00.4s $\pm$ 4.4s and $\delta$=$-63$ deg $11'55''$ $\pm 3''$ \cite{a1}): at least 5 catalog candidates are listed in several counterpart categories, from pulsars, to HII regions, to radio sources, to X-ray sources. If we were to determine some sort of FoM for each of these categories, how would them compare? Would a 95\% CL for the positive identification of a blazar as a counterpart be better than a 90\% CL for the positive identification of the source as an X-ray binary? There is no reliable to this,  given the fact that FoM for different populations are not directly comparable. They test relative goodness within a closed set (the class), and not an intrinsic property equally ascribable to any astrophysical object. This is an intrinsic limitation, given that the very process of assigning merit depends on the population considered. In addition, for the former example H.E.S.S. J1303-631 is an extended source and probably none of the positional coincident counterparts within its error localization is directly responsible for the $\gamma$-ray flux.}

{\it  The problem of time: } {\small  At present, the most successful identification scheme for $\gamma$-ray 
sources is based upon multifrequency follow-up observations
unless there is a given prediction of periodicity, which itself would 
unambiguously label the source. Simply put, the anticipated number of unidentified detections will preclude
making an individual multifrequency study for every
source, in the way it led to the identification of many
$\gamma$-ray blazars and the Geminga pulsar.}

\section{Identification of populations}

The ROSAT catalog of bright sources contains 18811 detections. Its extension, the faint catalog, contains 105924.\footnote{Both are available in the internet at http://www.xray.mpe.mpg.de/rosat/survey/ } Why are we not pursuing an identification process of one of such 
sources, yet unclassified?  Is it because we would find no a priori reason why to work on a particular 
source and not on any other of the thousands there are to choose? 
Or is it that we believe that what we will find after the study is one source id 
pertaining to one of the known X-ray classes (put otherwise: that we believe 
more or less understand the X-ray populations)? For GLAST, beyond identifying individual detections, there is a need to 
develop the tools that will make us discover first the populations for 
which LAT sensitivity allows.

One approach for identifying $\gamma$-ray populations in the LAT catalog was advanced in \cite{aabb}. Not emphasizing here on the details of the concept and management of budget probabilities, it proposes a protocol for population search with three steps, 1) a  theoretical selection: to define a priori of the unblinding of data which populations to search 2) a discovery protection: to manage the testing budget in a limited sample and 3) a quality evaluation: to precise the level of confidence in a equal basis for all populations. The key idea behind this protocol is to test the null hypothesis, such that finding a population is ruling it out. The ruling out of the null hypothesis can be quantified using small number statistics. To grasp at it, 
let ${\cal N}(A)$ be the number of known sources in the particular
candidate population $A$ under analysis and ${\cal U}$ the number
of LAT detections.
Let ${\cal P}$ be the probability that in a random direction of
the sky we find a LAT source.
%
The value of ${\cal P}$ is to be obtained
a priori of checking for any population.
The number of excess detections will be, ${\cal
E}(A)={\cal C}(A)-{\cal P} \times {\cal N}(A)$.\footnote{Notation: 
$P_A$ part of the probability budget a priori assigned to population $A$, such that to claim detection, $P^{\rm LAT}(A)<P_A$. 
$P^{\rm LAT}(A)$ is the random probability level by which we have found the population in the LAT
catalog. ${\cal C}(A)$ is the number of coincidences between members of population $A$ and LAT sources found.}
Let us consider the testing of a null
hypothesis (e.g., X-ray binaries are not LAT sources). That is
represented by 0 predicted signal events (coincidences), i.e.
total number of events equal to the background in Table 2-9 (see
leftmost columns) of \cite{FC}. Suppose for
definiteness that ${\cal P} \sim 3 \times 10^{-3}$ and ${\cal
N}(A)$ is equal to 200, then the number of chance
coincidences (the noise or background) is 0.5. Thus, if we find
more than 5 individual members of this class (e.g. superseding the
confidence interval 0.00-4.64) correlated with LAT sources, we
have proven that the null hypothesis is ruled out at the the 95\%
CL.
Using small number statistics we can convert the
level of confidence achieved for each population into the factual
probability, i.e., $P^{\rm LAT}({\rm X-ray\,bin})$. Subsequently,
by comparing with the a priori budgeted requirement (i.e., is
$P^{\rm LAT}({\rm X-ray\, bin.}) < P_{\rm X-ray\, bin.}$?, we will
be able to tell whether the population has been discovered.
Managing $P_A$ is equivalent to requesting different populations
to appear with different levels of
confidence.

\section{GLAST-LAT and the new populations}

In this section, we provide a few comments (though with  no claim of being comprehensive) on key features of some of the expected new populations of $\gamma$-ray sources.

{\it SNRs/PWN:}
We know at least 19 positional coincidences between EGRET sources and SNRs [their molecular environment was
studied in all cases  and it is particularly notable for cases such as W28, IC443, and others \cite{TR} and references therein]. There could be the chance for LAT to find 
multiple sources from the same accelerator, entertaining the possibility to study diffusion of cosmic rays, although LAT sensitivity could not be enough for understanding an energy-dependent diffusion coefficient (e.g, \cite{123}.
GeV $\gamma$-ray imaging might directly resolve acceleration sites, but probably with less resolution than ground based experiments with larger collection area. 
$\gamma$-ray spectroscopy (GeV to TeV) distinguishes p from e origin at least for some cases (see Funk, these proceedings).
Several pulsar wind nebulae are now seen with H.E.S.S., and interpreted as inverse Compton emission from CMB, starlight, and IR light from dust. Key features in the modeling of spectrum (spectral peak + overlap with synchrotron) are for GLAST to test (e.g., Aharonian et al. 2006). 

 {\it Colliding winds of massive binaries} have long been considered as potential sites of non-thermal high-energy photon production. Both leptonic (inverse Compton of relativistic electrons with the dense photospheric stellar radiation fields in the wind-wind collision zone, e.g., \cite{br,reim}) and hadronic (neutral pion decay products, where mesons produced by inelastic interactions of relativistic nucleons with the wind material produce the $\gamma$-rays, e.g., \cite{Ben,reim}) scenarios have been developed. Additionally, inverse Compton pair cascades
initiated by high-energy neutral pion decay photons (from nucleon-nucleon interactions in the stellar winds, e.g., \cite{B05}) or collective wind scenario in young stellar clusters (e.g., diffusive shock acceleration by encountering multiple shocks, e.g., \cite{Kle};
or inelastic proton interactions with collective winds taking into account the possible convection of cosmic ray primaries, e.g., \cite{Tor,DT06} have been also put forward. Predictions for the LAT have been developed in most of these models, and are subject to testing. The recent detection of Wd2 by H.E.S.S. gives confidence that these class of sources can energize particles well beyond the LAT domain. 

 {\it Microquasars and X-ray binaries:} After the detection of LS 5039 and LSI +61 303 \cite{a2,aa}, there is little doubt that the correspondingly coincident EGRET sources are physically connected. LAT will directly test this, and feedback the many recent theoretical models for these sources: leptonic (e.g., \cite{BBB,Gup}), hadronic (e.g., \cite{retal}), those involving cascading process (e.g,  \cite{be06}), etc. The possible interaction of jets with the ISM at larger scales will also be tested 
(e.g., \cite{BBBBB}).

 {\it Molecular clouds}, especially those that are close and at high $b$ (even with no accelerator nearby) could be $\gamma$-ray sources, some extended. There are pproximately 150 such clouds at $|b| >$ 10, most at about 100 pc-scales \cite{Tor-cloud}. Would LAT be able to 
map in $\gamma$-rays the nearby ISM? Would it be able to track the different cosmic ray distributions in them? The former answer is probably yes, the latter depends on how well we manage the uncertainty in the X factor and distance to the clouds.

{\it Galaxies: Normal galaxies to ULIRGs:}
None of them have been detected so far (for the observational status see \cite{T004,Cillis}, but 
many have solid predictions already on the verge of EGRET detectability \cite{V,P,DT05,T04}.

 {\it Galaxy clusters:} Multifrequency observations confirm non-thermal activity (non-thermal X-rays, diffuse radio halos) in these systems, and the 
hierarchical merging  implies Fermi 1st order acceleration and turbulence.
None have been detected so far, for the current observational status see \cite{RE}).

\section{Concluding remarks}

There is a need to apply a method of classification and identification of classes of $\gamma$-ray sources besides individuals. There are difficulties in assessing population identifications with FoM generally defined within a 
particular class. The systematics problems from number of sources and not only limited to the size of error localization boxes. There is however good reasons to think that LAT will identify many new populations, at 
least many key astrophysical scenarios have solid predictions above GLAST sensitivity, and several of them are already convincingly
identified at higher energies.

{\it Acknowledgements:}
{\small Work supported by the Ministery of Education of Spain,  grant PNAYA 2006-00530 and the Guggenheim Foundation. The author acknowledges O. Reimer for discussions.}


\begin{thebibliography}{99}


\bibitem{123} Aharonian F. \& Atoyan A. 1996, A\&A 309, 917

\bibitem{a1} Aharonian et al. (H.E.S.S. Collaboration) 2005a , A\&A 439, 1013
\bibitem{a2} Aharonian F. A. et al.  (H.E.S.S. Collaboration) 2005b, Science 309, 746
\bibitem{a3} Aharonian F. et al. (H.E.S.S. Collaboration) 2006, A\&A 448, L43

\bibitem{aa} Albert J. et al.  (MAGIC Collaboration) 2006, Science 312, 1771

\bibitem{B05} Bednarek W. 2005, MNRAS 363, L46
\bibitem{be06} Bednarek W. 2006, astro-ph/0611291, to appear in A\&A

\bibitem{Ben}  Benaglia P., et al. 2001 A\&A
366, 605

\bibitem{br} Benaglia P., \& Romero G. 2003, A\&A 399, 1121

\bibitem{BBB} Bosch-Ramon V., Romero G. E. \& Paredes J. M. 2006, A\&A 447, 263
\bibitem{BBBBB} Bosch-Ramon V., Aharonian F. A. \& Paredes J. M. 2005, A\&A 432, 609

\bibitem{Cillis} Cillis A. et al. 2005, ApJ 621, 139

\bibitem{DT05} Domingo-Santamar\'{\i}a E. \& Torres D. F. 2005, A\&A 444, 403
  
\bibitem{DT06} Domingo-Santamar\'{\i}a E. \& Torres D. F. 2006, A\&A 448, 613


\bibitem{FC} Feldman G. J. \& Cousins R. D. 1998, Physical Review D
57, 3873

\bibitem{Gup} Gupta S., Boettcher M., \& Dermer C. 2006 ApJ 644, 409

\bibitem{Kle} Klepach, E. G.; Ptuskin, V. S.; Zirakashvili, V. N. 2000, Astroparticle Physics 13,  161


\bibitem{la} La Palombara N. et al. 2006, astro-ph/0607251

\bibitem{P} Paglione, T. A. D., et al. 1996, ApJ, 460, 295

\bibitem{reim} Reimer A., Pohl M, \& Reimer O. 2006, ApJ 644, 1118

\bibitem{RE} Reimer O. et al. 2003, ApJ 588, 155
\bibitem{retal} Romero G. E. et al. 2003, A\&A 410, L1

\bibitem{rbt} Romero G., Benaglia P., \& Torres D. F 1999, A\&A 348, 868

\bibitem{TR} Torres D. F. et al. 2003, Phys. Rev. 382, 303
  
\bibitem{aabb} Torres D. F. \& Reimer O. 2005, ApJ 629, L141
 
 \bibitem{Tor-cloud} Torres D. F., Dame T. M. \& Digel S. W. 2005, ApJ 621, L29
   
\bibitem{Tor} Torres D. F., Domingo-Santamaria E., \& Romero G. E. 2004a, ApJ 601, L75
 
 
\bibitem{T004} Torres D. F. et al. 2004b, ApJ 2004, 607, L99
  
\bibitem{T04} Torres D. F. 2004, ApJ 617, 966

\bibitem{V} V\"olk H., Aharonian F. A. \&
Breitschwerdt D. 1996, Space Science Reviews 75, 279


\end{thebibliography}
\end{document}